\documentclass[review]{elsarticle}
\graphicspath{ {./figures/} }
\usepackage{hyperref}
\usepackage{float}
\usepackage{verbatim} 
\usepackage{apalike}
\restylefloat{figure}
\restylefloat{table}
\usepackage{lscape}
\usepackage{array}
\usepackage{booktabs}
\usepackage{csvsimple}
\usepackage{graphicx}  
\usepackage{rotating}
\usepackage{caption}
\usepackage{longtable}
\usepackage{ltxtable}
\usepackage{etoolbox}
\usepackage{algorithm}
\usepackage{algorithmic}
\usepackage{adjustbox}

\journal{Expert Systems with Applications}

\biboptions{authoryear}

\begin{document}
\begin{frontmatter}

\begin{titlepage}
\begin{center}
\vspace*{1cm}

\textbf{ \large Using a Deep Learning Model to Simulate Human Stock Traders' Methods of Chart Analysis}

\vspace{1.5cm}

Sungwoo Kang$^{a}$ (sungwoo919@hanmail.net), Jong-Kook Kim$^b$ (jongkook@korea.ac.kr)

\hspace{10pt}

\begin{flushleft}
\small  
$^a$ Department of Electrical and Computer Engineering, of Korea University Seoul 02841, Republic of Korea \\
$^b$ School of Electrical Engineering, of Korea University Seoul 02841, Republic of Korea \\


\end{flushleft}        
\end{center}
\end{titlepage}

\title{Using a Deep Learning Model to Simulate Human Stock Traders' Methods of Chart Analysis}

\author[1]{Sungwoo Kang}
\ead{sungwoo919@hanmail.net}

\author[2]{Jong-Kook Kim: corresponding author}
\ead{jongkook@korea.ac.kr}

\begin{abstract}
Despite the efficient market hypothesis, many studies suggest the existence of inefficiencies in the stock market leading to the development of techniques to gain above-market returns. Systematic trading has undergone significant advances in recent decades with deep learning schemes emerging as a powerful tool for analyzing and predicting market behavior. In this paper, a method is proposed that is inspired by how professional technical analysts trade. This scheme looks at stock prices of the previous 600 days and predicts whether the stock price will rise or fall 10\% or 20\% within the next D days. The proposed method uses the Resnet's (a deep learning model) skip connections and logits to increase the probability of the prediction. The model was trained and tested using historical data from both the Korea and US stock markets. The backtest is done using the data from 2020 to 2022. Using the proposed method for the Korea market it gave return of 75.36\% having Sharpe ratio of 1.57, which far exceeds the market return by 36\% and 0.61, respectively. On the US market it gives total return of 27.17\% with Sharpe ratio of 0.61, which outperforms other benchmarks such as NASDAQ, S\&P500, DOW JONES index by 17.69\% and 0.27, respectively.

\end{abstract}


\end{frontmatter}

\section{Introduction}
Stock market prediction has always presented challenges due to its complexity and unpredictable nature. It was thought that it is impossible to make an above-the-market profit because of the efficient market hypothesis. However it is not the case and research show that there's some inefficiency in the market \cite{jegadeesh1993returns,asness2013value,fama2008dissecting}, and many techniques have been developed to achieve this marginal gain, called the alpha. 
Reports such as \cite{nyfunds} show that making above-the-market profit is difficult. Over the last 20 years, less than 10\% of actively managed funds in the U.S were shown to beat the market, and no mutual fund was shown to beat the market during the five years between 2017 and 2021.

Systematic trading has undergone significant advances in recent decades, with new approaches transforming the field. 
In earlier days, traders employed basic technical methods such as moving averages, Bollinger Bands \cite{bollinger2002bollinger}, and Relative Strength Index (RSI) \cite{arksey2021rsi} to analyze market trends. These methods evolved to strategies such as taking long positions on high-percentile stocks and short positions on low-percentile ones, based on factors such as momentum, volatility, and growth. In addition to these approaches, attempts were made to predict stock prices through time series analysis techniques like ARIMA (auto-regressive integrated moving average) \cite{Harvey1990}. 

Recently, deep learning has emerged as a powerful tool for analyzing and predicting the market behavior by learning complex patterns and relationships in financial data. Many different methods are used to predict the market behavior.
In the belief that social media precedes future events, attempts to predict stock price by analyzing social media and combining Natural Language Processing exist \cite{9006342,maqsood2020local,xu2019stock}. Because the stock price is a time series, approaches using Long Short-Term Memory models (LSTM)s \cite{zou2020using,ghosh2022forecasting,fabbri2018dow} have been attempted. \cite{wu2020adaptive,liu2018practical,yang2020deep} proposed Reinforcement Learning Algorithms to discover trading rules. Approaches using time series values as CNN inputs (\cite{gudelek2017deep, takeuchi2013applying}) and using an image of the stock chart as data input have been studied. \cite{kusuma2019using}.

Professional technical analysts somehow make a consistent profit without these sophisticated approaches. By just looking at the charts, these human traders find major and minor trends as well as support and resistance levels of the stock. Support and resistance levels are price levels in trading where buying (support) or selling (resistance) pressure is expected to be significant, often leading to price reversals or consolidation. Inspired by these traders, this paper attempts to simulate their approach using a deep learning model. The underlying premise is that there are certain price patterns that occur before a stock rises in the short term. The model utilizes historical stock prices over a period of 600 days to predict whether there will be a 10\% rise or fall in the stock price within the next D days. We assume that there are specific spanning time that this patterns occur, and try to find them. 
The Resnet model that was applied to a time series problem \cite{wang2017time} is used in this paper.
Although one might suggest LSTM is more appropriate for time series problems, we think Resnet is suit for the task because these traders trade by drawing trend lines on charts (which is a 2D image). For these traders, finding out short-term and long-term trends is important when trading. 
Using the Resnet's skip connections, the model is expected to capture these major and minor trends. For example, using a kernel size of 5 would extract features similar to the 5-day moving average in the first layer, and the next layer will again calculate the 5-day average with this first layer's output which results in extracting longer-term features. When trading, the model will buy stocks that are expected to rise by 10\% or 20\% in the next D days. We view this as classification problem instead of time series regression because in stock market there are instances where a stock may rise but the timing of its rise is uncertain.
In order to increase the probability of success in stock trading, stocks whose largest softmax logit from the Resnet that exceeds a certain threshold are considered for trading. By setting this threshold, the trading is focused on stocks that the model predicts with higher confidence, potentially increasing the chances of achieving profitable trades. As far as the authors know, this paper is the first to utilize softmax logits of a deep learning model for stock trading. 
The model was trained using all tickers from KOSPI, KOSDAQ market for Korea stocks, and tickers from NYSE, NASDAQ, AMEX for US stocks from year 2006 to 2015, with validation period set to year 2016 to 2019.
The model was tested on historical data from year 2020 to 2022, giving a profit of more than 49\% above the Korea market return, and a profit of more than 40\% above the US market return. 

The success of our method can be attributed to the combination of these: 
\begin{itemize}
  \setlength{\itemindent}{0em}
  \item using Resnet model with a long window size of 600 days
  \item assigning labels to stock price data using whether the price will rise or fall by 10\% or stays between the rise and fall range in the given days of 1, 3, 5, 10, 15, 20, 30
  \item only trading stocks that have softmax logit value above a certain threshold chosen
\end{itemize}
The paper is organized as follows. In the first section, the motivation for the research is explained. Previous research and the differences from this research are described in the next section. The third section explains the approach proposed in this paper and the next section shows the results and discussions. Finally, the last section concludes the paper, and insights when using this technique are depicted.

\section{Related Work}
Using a sliding window and assigning labels to price movement with discretized return rate is not new. By discretizing the return rate, one can overcome small price change sensitivity and facilitate model training. \cite{HAN2023118581} discussed the effect of N-Period Min Max labeling instead of just predicting whether the price will go up or down and used the XGBoost model to trade accordingly. They assigned labels to stock price movement according to which side of the upper or lower price limit of the given period is hit first.
\cite{TEIXEIRA20106885} combined technical indicators with nearest neighbor classification on fifteen real stocks from Sao Paulo Stock Exchange. Since trading volume provides important information, \cite{CHAVARNAKUL20081004} combines equivolume charting with generalized regression neural network to predict S\&P 500 index. \cite{DASH201642} proposed computational efficient functional link artificial neural network (CEFLANN) with technical indicators to output trading signal for BSE SENSEX and S\&P 500 index. 
\cite{gudelek2017deep} used a 2-dimensional CNN to predict the next day's return of commonly used ETFs. 28 features with a 28-day window were used as input to the CNN which includes close price, volume, and several technical indicators such as Relative Strength Index (RSI), Simple Moving Average (SMA), Moving Average Convergence Divergence (MACD). \cite{takeuchi2013applying} used an autoencoder composed of stacked restricted Boltzmann machines. They used 12 monthly returns and 20 daily returns as features and predict whether the stock's return will be above or below the median return of whole stocks.  
\cite{mukherjee2023stock} trained a CNN network by inputting 2-D histograms generated out of the quantized dataset within a particular time frame. It was tested during the recent COVID-19 pandemic and achieved an accuracy of 97.66\%. 
Our approach differs from these approaches in that only OHLCV (open, high, low, close, volume) price data was used without any human-made features, uses a long window of 600 days to capture price trends, and assign labels by whether the price hits a fixed return rate of 10\% for a given period. Sample weighted loss function was also used to improve class imbalance and utilize softmax logits for better accuracy and cumulative return. Backtest results on 2020-2022 is also provided that significantly outperform the market. Providing backtest results is important because high prediction accuracy does not lead to long-term profits if the Profit and Loss (P/L) ratio is poor.

\section{Proposed Approach}
\subsection{Dataset Preprocessing}
OHLCV (open, high, low, close, volume) information of daily stock prices of all the tickers (KOSPI, KOSDAQ tickers in Korea market and NYSE, NASDAQ, AMEX tickers in US) are collected from Naver Finance (finance.naver.com/) for Korea stocks, and historical data from Stooq (www.stooq.com) is used for US stocks. The log price (base 10) was used instead of daily returns because we suspect the stock price range provides important information and using the original price will lead to exploding gradient. 
The label was determined by observing the stock price for the preceding 600 days and recording whether the stock price increases or decreases by 10\% within the next D days, compared to the closing price. We believe that discretizing returns is important to facilitate learning and avoid overfitting. 10\% was chosen because the Volatility Interrupt (VI), a mechanism that pauses the market for a specified period of time to allow market participants to reassess the situation and prevent excessive price swings, is activated in Korea market when the price rises by 10\%. More than 10 tickers rise by 10\% every day, although this number differs largely according to market conditions. Label for 20\% was made also. 
The model would only see these OHLCV prices of 600 days (input dimension of 5*600) and would have no information of which ticker it is extracted from or any other fundamental factors such as corporate earnings, dividends, and cash flows. The model would only see the 'chart patterns' to predict the future price. 
The window of 600 days is chosen to encompass the effect of 480 day moving average. In human experts' technical analysis, the moving average is a tool used to smooth out price data over a specific time period. When the stock's price approaches this long-term moving average (such as 480-day moving average), it can act as a support level that doesn't let the stock price to fall below. 
The variable D was experimented with different values of 1, 3, 5, 10, 15, 20, and 30 to evaluate their impact on the model's performance. 
In the 5-day 10\% scheme, a label of 2 is given if the highest price of the next 5 days is 10\% or higher, 1 if both the highest and lowest prices are in the range of -10\% to 10\%, and 0 if the lowest price is less than -10\%, during the next 5 days.
Note that the 5-day 10\% scheme and 10-day 10\% scheme will have the same label of 2 if the price reaches 10\% on the third day and falls -10\% on fourth day. 
If both the highest and lowest prices reached the set limit of +/-10\% on the same day, they were treated as label 2 although we don't know which limit the stock price reached first. If the lowest price reached the -10\% first the trade would result in loss, while if the highest price reached +10\% first it would result in gain in real trade. 

A well-chosen back-testing period is important to avoid over-fitting. 
In general, a simple buy\&hold strategy is advantageous during a bullish market (when stocks prices tend to keep increasing). On the other hand, in a sideways market where prices are moving within a range, it is advantageous to trade when the price touches either end of the range. Strategies optimized specifically for either upside or sideways markets may not work as effectively in different market conditions. Hence training and testing periods were carefully chosen to avoid this. The training period was chosen to be 2006-2015, and the validation period to be 2016-2019, both of which include strong bear and bull markets and markets that go sideways. The model was tested from 2020 to 2022. For US dataset, tickers that have close prices above \$2000 and below \$2 were excluded in training, as stock prices range widely from penny stocks to above \$50,0000. Korea stock dataset consists of 2673488 training samples, 1611189 validation samples, and 1477200 test samples while US stock dataset consists of 3807869 training samples, 2409155 validation samples, and 2369288 test samples. 

Dataset statistics for Korea and US stocks are shown in table \ref{table:kor_dataset} and \ref{table:us_dataset}. Each item in the table shows a percentage of 10\% or 20\% rise, fall, and the stocks that went sideways in this range during the given period rounded to second decimal separated by '/'. For example, the training dataset of US with labeling period 3 shows 0.07/0.07/0.86, which means that during the period of 3 days, roughly
7\% of stocks reached a 10\% price rise, 7\% of stocks reached 10\% price fall, and 86\% stayed in the price range of -10\%~10\% fall during 2006-2015. 
The 'uncertain' columns shows the percentage of the samples classified as 10\% rise although we don't know because the price fluctuated twice the given label percentage. However, to adopt a more conservative stance during backtesting, these instances are treated as losses.
It can be seen that the longer the labeling period, the 10\% rise or fall ratio gets higher and the sideways ratio gets lower. This can be interpreted as stocks that move sideways determining direction over time.

\begin{table}[t]
\centering
\csvautobooktabular{merged_naver_stat.csv}
\caption{dataset statistics for Korea stocks}
\label{table:kor_dataset}
\end{table}

\begin{table}[t]
\centering
\csvautobooktabular{merged_stooq_stat.csv}
\caption{dataset statistics for US stocks}
\label{table:us_dataset}
\end{table}

\subsection{Proposed Model}
\begin{figure}[!htbp]
\centering
   \includegraphics[width=8.5cm]{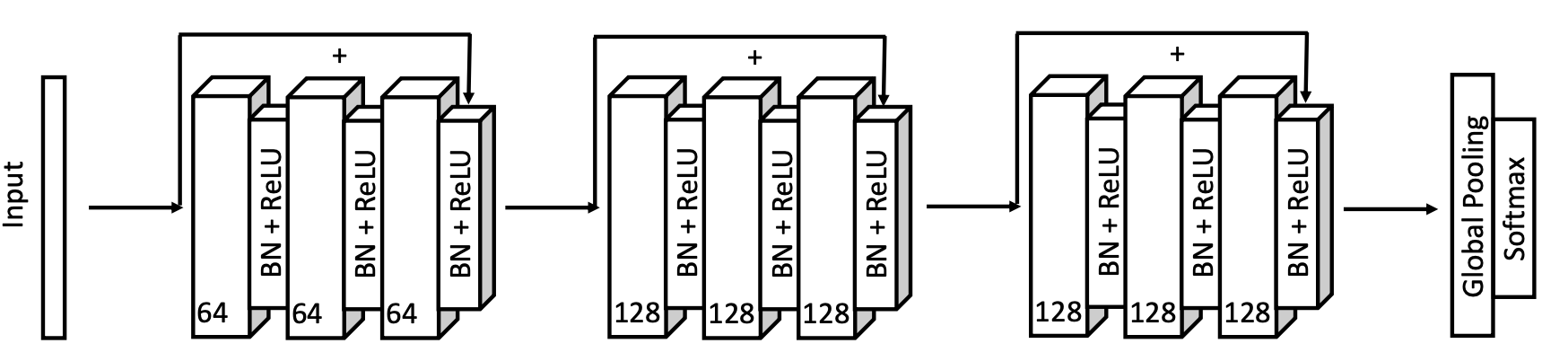}
   \hfil
\caption{Model architecture of \cite{wang2017time}}
\label{tsai_resnet}
\end{figure}
\begin{figure}[!htbp]
\centering
   \includegraphics[width=8.5cm]{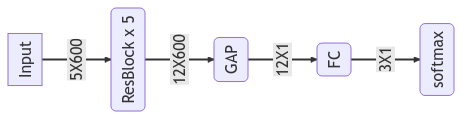}
   \hfil
\caption{the modified model architecture used in this paper}
\label{our_model}
\end{figure}

ResNet \cite{he2016deep} is a deep learning model architecture designed to overcome the vanishing gradient problem by incorporating skip connections. These connections enable the network to retain important information from earlier layers, making it easier to train deeper neural networks. \cite{wang2017time} modified Resnet's architecture to be used in time series. This model is modified for training. 
The Resnet architecture's skip connections is expected to capture major and minor trends, lower layers to capture short-term trends, and higher layers the long-term. For example, using a kernel size of 5 would extract features similar to the 5-day moving average in the first layer, and the next layer will again calculate the 5-day average with this first layer's output which results in extracting longer-term features. This process continues through the subsequent layers. And because skip connection is used, this information would be gradually integrated. 
The architecture of \cite{wang2017time} is modified in this application. The architecture of \cite{wang2017time} and the model proposed in this paper is shown in \ref{tsai_resnet} and \ref{our_model}. The model gets an input dimension of 5x600, representing the log OHLCV price of 600 stock days. A one-dimensional Resnet model with 5 blocks with a fixed channel number of 12 is utilized, then passed to the Global Average Pooling (GAP) layer to get the mean values of each feature map. Each block consists of 3 convolutional layers with kernel sizes of 7,5,3 each with Batch Normalization (BN) and ReLu activation applied. A skip connection is added after each block. The number of channels is fixed to 12 all thorough. After passing the Fully Connected Layer softmax values were calculated representing the probability of a 10\% (20\%) rise, 10\% (20\%) fall, and price between -10\% to 10\% (-230\% to 20\%). The Binary Cross Entropy (BCE) is used as the loss function. When implementing the model, Resnet implementation for time series in this paper \cite{wang2017time} is used. The model is trained with kernel stride of 1, padding of 1. All models were trained for 50 epochs, with 16-bit precision. Training took 1.5 days for Korea stocks and 5 days for US stocks. 
The same model using varying kernel sizes 5,15,30,60,120 and 5,15,30,60,120 was also trained. The kernel size is same within block with these models. These varying kernel sizes were expected to capture various short and long-term trends but were shown to be ineffective compared to fixed kernel size of 7,5,3. 

\subsection{Choosing the Best Model and Threshold}
The softmax logit (output) of a model is commonly believed to indicate the level of confidence in the result of an image recognition task, such as identifying 90\% happiness and 5\% surprise in emotion recognition. 
We believe this is also the case for trading. By only predicingt for which the largest softmax value of the model’s last layer (logit) is above a certain threshold, we can increase model profitability. We refer this as ‘threshodling’. This would not be possible for normal classification problem as we have to give result for all the instances, but in trading we can only confident stocks.


Apart from improving the probability of success, there is another reason to use thresholding in trading. There are around 4000 (Korea) and 5000 (US) tradable tickers available. Even trading 1\% of the dataset would still result in about 40-50 tickers daily, which is not practical and would incur hefty fees. Establishing a threshold value narrows down the list of tradable stocks, thus reducing both the number of trades and the fees associated with them.

We show part of the validation result table for model trained on US market with various thresholds on table \ref{table:us_val_selected}. Accuracy, f1 macro score, precision and recall for each label name, threshold, and regarding dataset proportion is shown. All numbers are rounded up to four decimals. Threshold of 0, 0.7, 0.8, 0.9, 0.99, 0.999, 0.9995 are tested. Each statistics are calculated only with the items that are output by model above the given threshold. 'dataset proportion' denotes what fraction of its original dataset the metrics were calculated. 
Precision is the ratio tp / (tp + fp) where tp is the number of true positives and fp the number of false positives. Recall is the ratio tp / (tp + fn) which is a measure of the ability of a classification model to capture and correctly identify all relevant instances. 'f1 macro' denotes the F1 score with macro averaging. It calculates F1 score for each label and averages them. F1 score is the harmonic mean of precision and recall. 'nan' represents not a number since no instances were outputted to calculate those scores. 

\begin{table}[!htbp]
\centering
\resizebox{\linewidth}{!}{%
    \csvautobooktabular{val_us_accuracy_result_selected.csv}
}
\caption{Part of Validation set statistics for US label}
\label{table:us_val_selected}
\end{table}



It can be seen that as you increase the threshold, the overall accuracy increases while the corresponding proportion of dataset decreases. Because the model only predicts confident instances, the accuracy reaches almost 1 (100\%) for some labels with dataset proportion close to 0 (0\%).
Some models seems to fail in the learning process, specializing in predicting specific labels. For example, US model label\_1\_tp20\_ls20 only predict for label 1 above threshold 0.8, which gives 1 in label 1 precision but all others 0. In US model label\_20\_tp20\_ls20, the precision of label 1 increase as you increase the threshold but precision of label 0 and 2 decreases as you increase the threshold. We think this problem arises due to dataset imbalance.

Selecting the appropriate label and threshold for backtesting is a problem since testing multiple combinations will only expose us to multiple comparison bias. To choose the best model, we hypothesize that successful model should be able to sufficiently predict all three labels (up, down, sideways) to learn the important features. 
We rank models by their F1 macro scores and choose the threshold that maximizes the score for each label. Subsequently, we only take the label and threshold which the respective dataset proportion exceeds 0.00001 to ensure an adequate number of trades are made. 
The according label name with respective threshold, dataset proportion, f1 score, and the accuracy ordered by f1 score is shown on table \ref{table:kr_val_bt_labels} for Korea market\ref{table:us_val_bt_labels} for US market. We proceed to backtesting with these labels. 

\begin{table}[!htbp]
\centering
\csvautobooktabular{us_val_label_to_backtest.csv}
\caption{labels chosen to be backtested on US}
\label{table:us_val_bt_labels}
\end{table}

\begin{table}[!htbp]
\centering
\csvautobooktabular{kr_val_label_to_backtest.csv}
\caption{labels chosen to be backtested on Korea}
\label{table:kr_val_bt_labels}
\end{table}

When using the model for trading, the model looks all the tickers at the end of the market every day, and buys the stock if the model predicts label 1 and the respective softmax logit is above threshold. Stocks bought are automatically sold if it increases or decreases 10\% or 20\%, or sold if N days passed according to trained label. The stock buying process using the model is illustrated in algorithm \ref{algo}

\begin{algorithm}
\caption{Stock selecting scheme using trained model}
\begin{algorithmic}[1]
\STATE Initialize model parameters and threshold
\FOR{each trading day}
    \FOR{each ticker}
        \STATE Retrieve 600 OHLCV price data for the ticker
        \STATE Feed data into the model
        \STATE Obtain predicted class and largest logit
        \IF{largest logit $>$ threshold AND predicted class = 1}
            \STATE Execute buy order for the stock
        \ENDIF
    \ENDFOR
\ENDFOR
\label{algo}
\end{algorithmic}
\end{algorithm}

In order to choose which model with the given period label performs the best, backtesting is conducted during the validation period. 
Original starting cash was set to be 10 million won for Korea stocks and ten thousand dollars for US stocks. Trading commissions were set to be 0.015\% every time trade is made and tax were set to be 0.20\% which is applied whenever the trade ended with a profit. We assume that the whole amount can be sold when the price reaches the given percentage, hence only +/-10\% or +/-20\% of P\&L (Profit and Loss) will occur during single trade. The entry ratio was chosen to be 1/20 of available cash. Although other entry scheme such as using fixed entry amount of 1/10 of original cash (100\$ per trade for US) and double it when the original cash doubles, this scheme was chosen to reduce the variance of each runs and compare with the index. 
Top ten results when ordered by Sharpe ratio is shown on table \ref{table:val_kr_bt_result} and \ref{table:val_us_bt_result}.

\begin{table}[!htbp]
\centering
\scriptsize
\csvautobooktabular{val_kr_bt_result.csv}
\caption{Korea market validation period backtest result top 10 ordered by Sharpe ratio}
\label{table:val_kr_bt_result}
\end{table}

\begin{table}[!htbp]
\centering
\scriptsize
\csvautobooktabular{val_us_bt_result.csv}
\caption{US market validation period backtest result top 10 ordered by Sharpe ratio}
\label{table:val_us_bt_result}
\end{table}

After the labeling period, a postfix '\_sidecut\_True' is assigned if the stock was sold at the closing price on days when it went sideways during the labeling period. Alternatively, postfix '\_sidecut\_False' is assigned if the stock was held even after the labeling period until it reached a given percentage of +/-10\%. For example, label\_5\_tp10\_ls10\_threshold\_0.99\_sidecut\_True refers to the model trained for a labeling period of 5 days, backtested with threshold 0.99, with entry ratio 0.1, and sells at the close price if the stock went sideways during the labeling period. In the real trading scenario, a 10\% drop followed by a 10\% increase will result in a loss, and vice versa as a gain. Because minute data for all tickers are not available in this research, it is unclear whether profit was made if the price fluctuated twice the labeling percentage on that day. 
'Trades' shows the total number of trades executed within the given period. Buying and selling one ticker counts as one trade. 'Return' shows the total return generated with respect to original cash during the test period. 'MDD' represents maximum draw-down is the maximum percentage decline in the trading account from its peak value to its lowest point. It is a measure that indicates largest unrealized loss the investor has to suffer during the trading period. 'Sharpe Ratio' is a risk-adjusted measure of return, the average return earned in excess of the risk-free rate per unit of volatility. The shapre ratio is used to compare various strategies since total return can be magnified with increased volatility using leverage. Higher volatility also means higher loss as well as higher profit. 
We choose our best model to be model\_label\_20\_tp10\_ls10\_sidecut\_False for Korea market and model\_label\_30\_tp20\_ls20\_sidecut\_False for US market, which ranks the best far exceeding others by 0.36 and 0.16 on Sharpe ratio.

\begin{figure}[!htbp]
\centering
   \includegraphics[width=11.5cm]{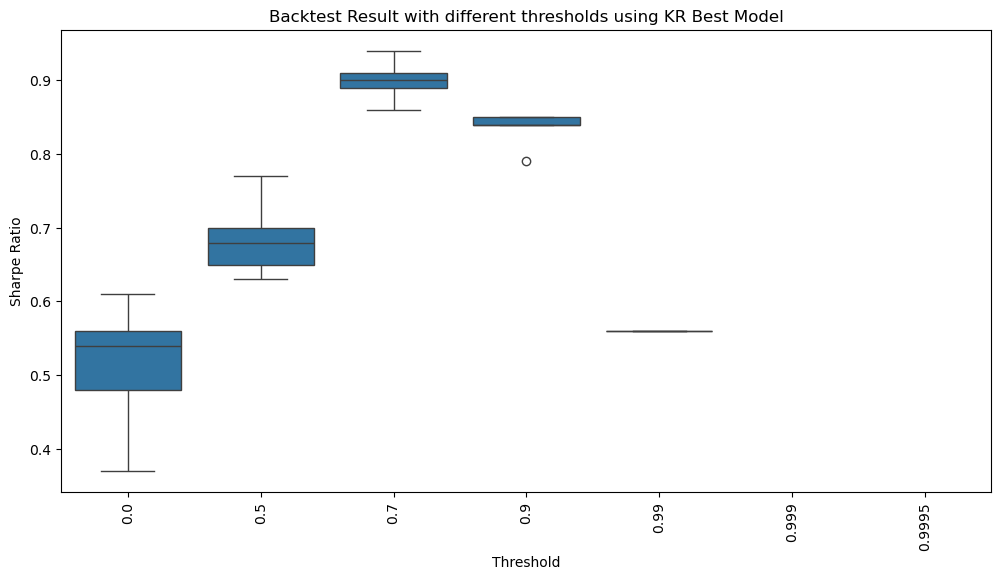}
   \hfil
\caption{Box Plot of Sharpe Ratio with different thresholds during validation period using KR Best Model}
\label{kr_threshold_plot_val}
\end{figure}

\begin{figure}[!htbp]
\centering
   \includegraphics[width=11.5cm]{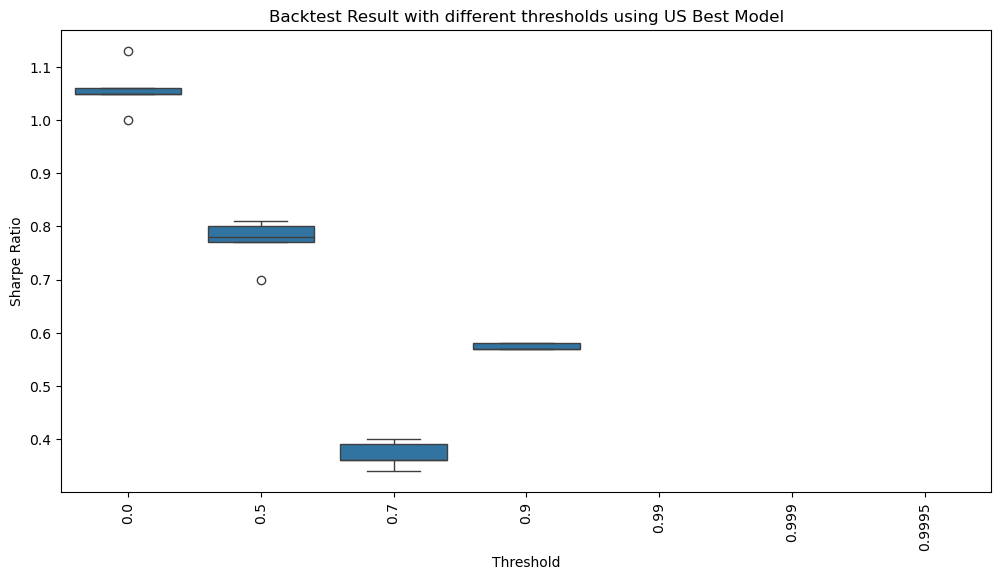}
   \hfil
\caption{Box Plot of Sharpe Ratio with different thresholds during validation period using US Best Model}
\label{us_threshold_plot_val}
\end{figure}

To ascertain the impact of threshold adjustments on model returns, we conducted backtesting on our top-performing model across a range of thresholds. As the threshold increases, both accuracy and F1 macro score tend to increase, but the proportion of the dataset that the model predicts decreases. We examined thresholds ranging from 0 (no threshold applied) to 0.9995. For each threshold, we executed five runs and analyzed the resulting Sharpe ratios. The boxplots illustrating these Sharpe ratios during the validation period for KR market is shown on Figure \ref{kr_threshold_plot_val} and and US market on Figure \ref{us_threshold_plot_val}. Threshold affects how frequently model will buy stocks as well as model accuracy. Because of this, we expected the Sharpe ratio plotted against the threshold to resemble a mountain shape, with the peak representing the optimal balance between accuracy and coverage. However for Korea market this was the case, for US market the Sharpe ratio tend to decrease as you increase the threshold. 
We believe this difference is likely due to market conditions. During the 4-year validation period, the Korean stock market only saw a maximum increase of 14.54\%, whereas the US S\&P 500 and NASDAQ rose by 60.52\% and 83\% respectively. In a strongly bullish market like the US during this period, holding a wider range of stocks including even less certain ones, would be advantageous. Even uncertain stocks are likely to end in profit. 


\section{Test Results}
\subsection{Comparison with Market Index}
The model is backtested on year 2020 to 2022 for test period. The Return (\%), maximum drawdown (\%), and Sharpe ratio of US and Korea's market index in shown on table \ref{table:market_stats}.
For Korea market, our best model (model\_label\_20\_tp10\_ls10\_sidecut\_False) far exceeds the total return by 72\% and the Sharpe ratio by 1.42.
However, for US market our best model (model\_label\_30\_tp10\_ls10\_sidecut\_False) fail to exceed total return and Sharpe ratio of the AMEX index although it gives 10\% more total return and 0.37 more Sharpe ratio than investing in S\&P500 and NASDAQ index. We think this is because the AMEX market performed extremely well during this test period. On validation period, AMEX ranks last compared to our model and other benchmarks in the validation period \ref{table:val_market_stats}. Our results far exceeds the Sharpe ratio of all the market index although it may just be a result of over-fitting.

\begin{table}[!htbp]
\centering
\scriptsize
\csvautobooktabular{market_stats.csv}
\caption{Return of Korea and US Market Indices during the Test Period}
\label{table:market_stats}
\end{table}

\begin{table}[!htbp]
\centering
\scriptsize
\csvautobooktabular{val_market_stats.csv}
\caption{Return of Korea and US Market Indices during the Validation Period}
\label{table:val_market_stats}
\end{table}

\subsection{Comparison of Quarterly Return of Market Index}
\begin{figure}[!htbp]
\centering
   \includegraphics[width=11.5cm]{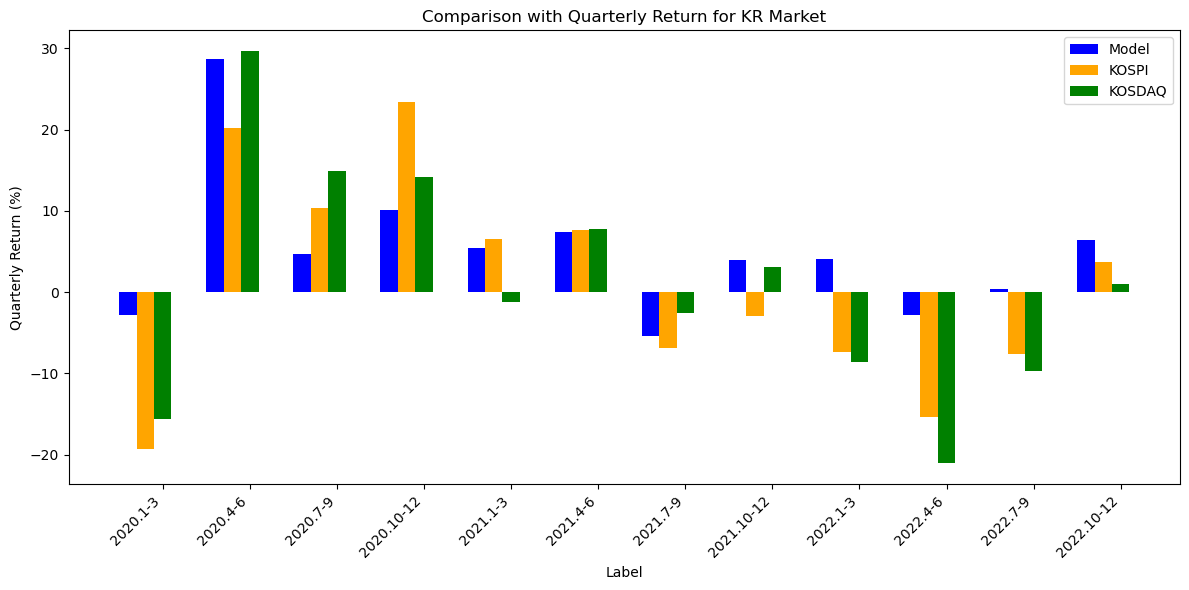}
   \hfil
\caption{Barplot of Quarterly return of KR market during the test period}
\label{kr_quarterly_barplot}
\end{figure}

\begin{figure}[!htbp]
\centering
   \includegraphics[width=11.5cm]{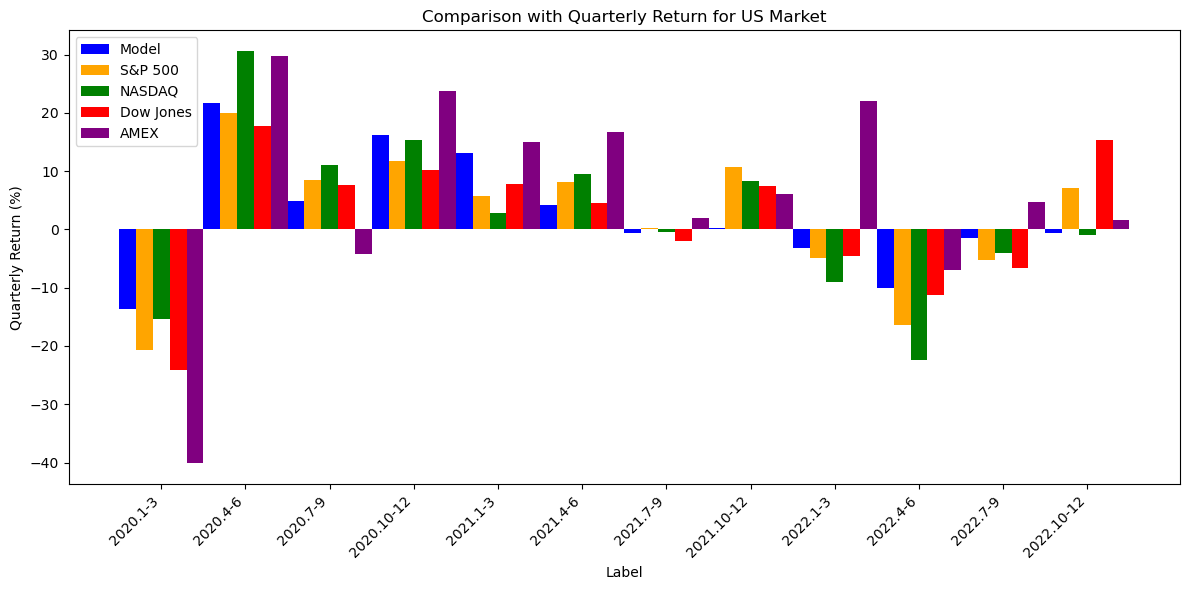}
   \hfil
\caption{Barplot of Quarterly return of US market during the test period}
\label{us_quarterly_barplot}
\end{figure}

We compare the quarterly return of our best model (model\_label\_20\_tp10\_ls10\_sidecut\_False for Korea market, and model\_label\_30\_tp10\_ls10\_sidecut\_False for US market) with the market return. The quarterly returns are calculated using one of the five random runs with the equity curve of the portfolio. The result table and bar plot of three month return during the test period is shown on table \ref{table:kr_quarterly}, \ref{table:us_quarterly} and figure \ref{us_quarterly_barplot}, \ref{kr_quarterly_barplot}. Although our model does not consistently beat the market on quarterly basis, big losses are minimized when they occur. 

\begin{table}[!htbp]
\centering
\csvautobooktabular{kr_quarterly.csv}
\caption{Quarterly return of KR market during the test period}
\label{table:kr_quarterly}
\end{table}

\begin{table}[!htbp]
\centering
\csvautobooktabular{us_quarterly.csv}
\caption{Quarterly return of US market during the test period}
\label{table:us_quarterly}
\end{table}

\subsection{Comparison with the Randomly Chosen Stocks and Market Index}
\begin{figure}[!htbp]
\centering
   \includegraphics[width=11.5cm]{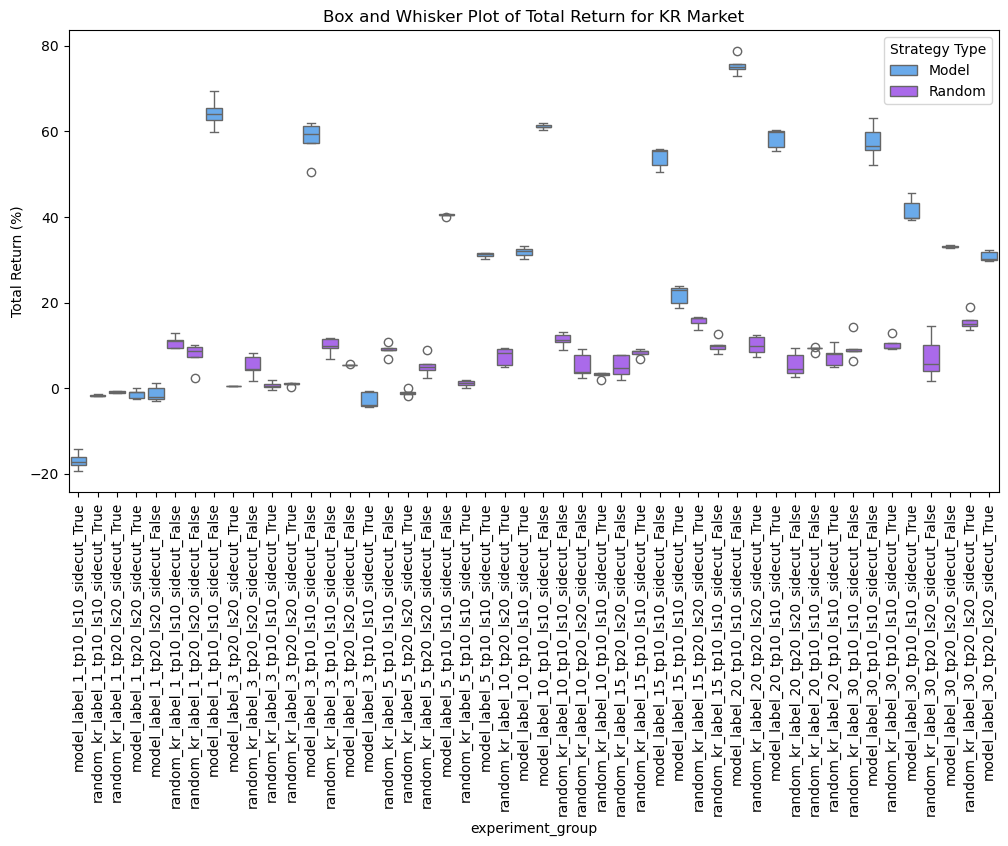}
   \hfil
\caption{total return of Korea market during the test period}
\label{kr_test_boxplot}
\end{figure}

\begin{figure}[!htbp]
\centering
   \includegraphics[width=12cm]{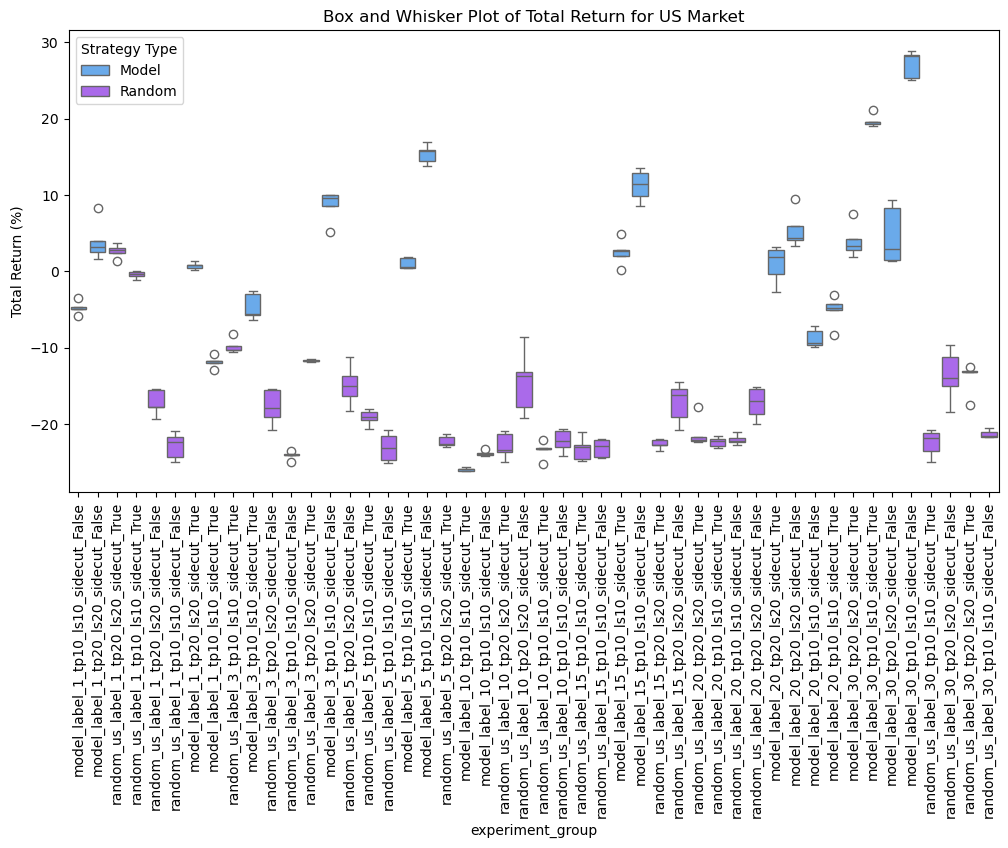}
   \hfil
\caption{total return of US market during the test period}
\label{us_test_boxplot}
\end{figure}

Our dataset includes survival bias, which means that stocks that are deslited are not included in historical prices. This can result in higher returns and Sharpe ratios, as surviving stocks are likely to have performed well. However preprocessed datasets that takes these into account are expensive. Hence, we compare the trained model with results of holding randomly selected stocks until they hit a predetermined percentage limit over the given period. These are shown with prefix 'random\_' in front of the label. Box plot of total return of five runs are shown on figure \ref{kr_test_boxplot}, \ref{us_test_boxplot}. Top ten result when ordered by Sharpe ratio is shown on table \ref{table:test_kr_bt_result}, \ref{table:test_us_bt_result}. If the model selects stocks that are most likely to rise, it should yield better returns than randomly selecting stocks excluding de-listed ones.
Despite the stocks are randomly selected, some number of backtests outperform the market return due to high turnover rate, which has the effect similar to investing higher amount of initial capital because when a stock is sold another stock is purchased immediately with the available cash.
Trained model generally tends to perform well than randomly chosen stocks, as indicated by blue label. 
For US market model\_label\_30\_tp20\_ls20\_sidecut\_False ranks the best consistent with the validation result. For Korean market model\_label\_20\_tp10\_ls10\_sidecut\_False ranks fourth when ordered by Sharpe ratio, however it gives the best total return and still far exceeds the randomly chosen market result. 
Although  gives the best total return on both valdation and test period,  when ordered by Sharpe ratio. However both results far exceeds the best random result.


\begin{table}[!htbp]
\centering
\scriptsize
\csvautobooktabular{test_kr_bt_result.csv}
\caption{KR market test period backtest result top 10 ordered by Sharpe ratio}
\label{table:test_kr_bt_result}
\end{table}


\begin{table}[!htbp]
\centering
\scriptsize
\csvautobooktabular{test_us_bt_result.csv}
\caption{US market test period backtest result top 10 ordered by Sharpe ratio}
\label{table:test_us_bt_result}
\end{table}

\subsection{Comparison with Other Work}
\begin{figure}[!htbp]
\centering
   \includegraphics[width=11.5cm]{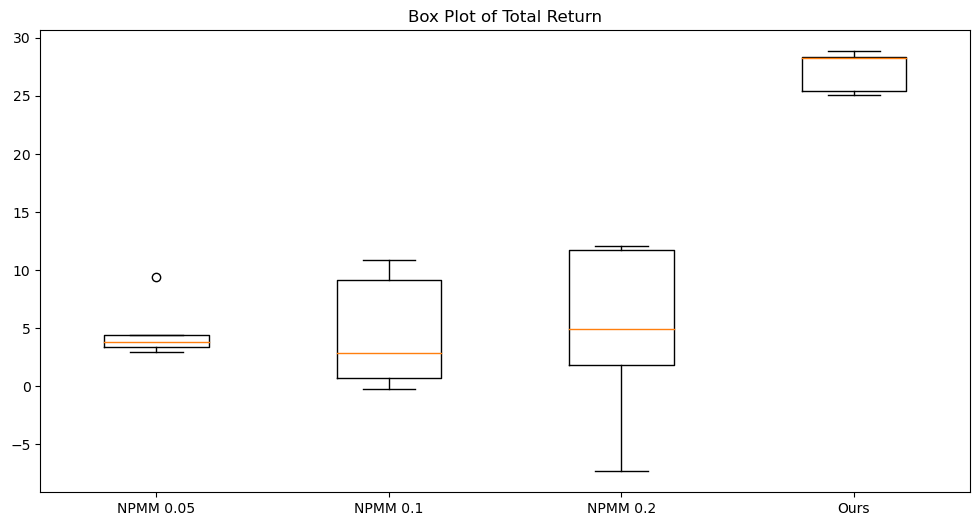}
   \hfil
\caption{total return of NPMM compared with Ours}
\label{npmm_return}
\end{figure}

\begin{figure}[!htbp]
\centering
   \includegraphics[width=11.5cm]{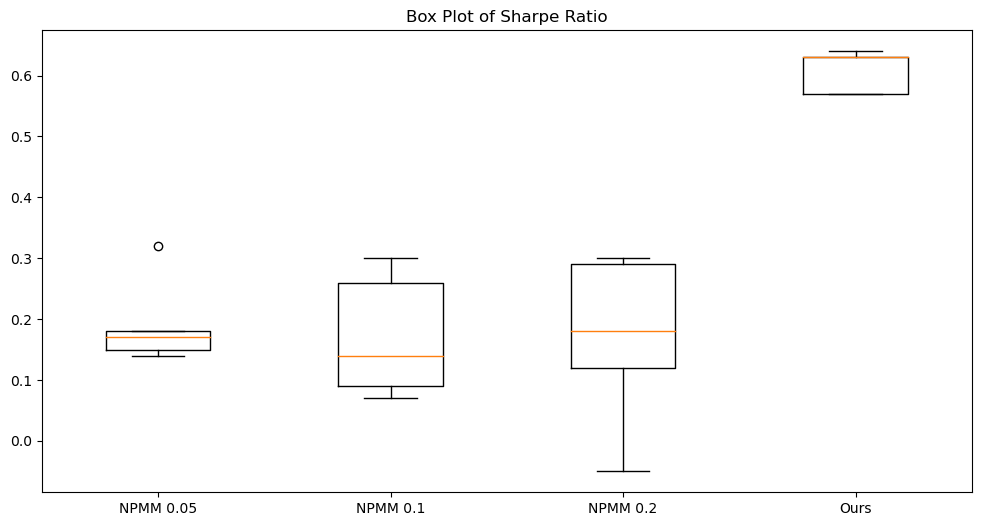}
   \hfil
\caption{Sharpe ratio of NPMM compared with Ours}
\label{npmm_sharpe}
\end{figure}

We compare other work with another labeling scheme on stock price. \cite{HAN2023118581} test the effect of N-period-min-max (NPMM) labeling with technical indiators and XGBoost algorithm. In their study, they assigned labels (0 or 1) based on whether the current price reached the lowest or highest price within a predefined window. While their model achieved profitability across 92 NASDAQ stocks, their evaluation focused on individual stock performance and did not address portfolio rebalancing strategies (what proportion of original cash to enter each trade).
When labeled with NPMM method very few points have labels,resulting in insufficient data points for XGBoost training on certain stocks. To compare with our result, 
we train XGBoost model on 84 tickers specified in the paper (8 tickers were excluded because data was not available) by setting t   he train period from 2009-01-02 to 2019-12-31. The model was backtested on 2020-01-02 to 2022-12-31. NPMM labeling period of 61 days was used, which was reported to have based performance in the paper. Entry ratio was tested with 0.05, 0.1, 0.2. 
The result which averages five runs are summarized on table \ref{table:npmm_bt_result}, and figures \ref{npmm_return}, \ref{npmm_sharpe}

\begin{table}[!htbp]
\centering
\csvautobooktabular{npmm_bt_result.csv}
\caption{NPMM Backtest Results compared with ours}
\label{table:npmm_bt_result}
\end{table}

When trading with NPMM labeling, only about 19 trades occurred giving return of 4.79\%, having Sharpe ratio of 0.19 and Maximum Drawdown of 17.35\%. Our best model far exceeds this with return of 35.54\%, having Sharpe ratio of 0.61 and Maximum Drawdown of 27.17\%.

\subsection{Effect of Thresholding on Backtest Result}
\begin{figure}[!htbp]
\centering
   \includegraphics[width=11.5cm]{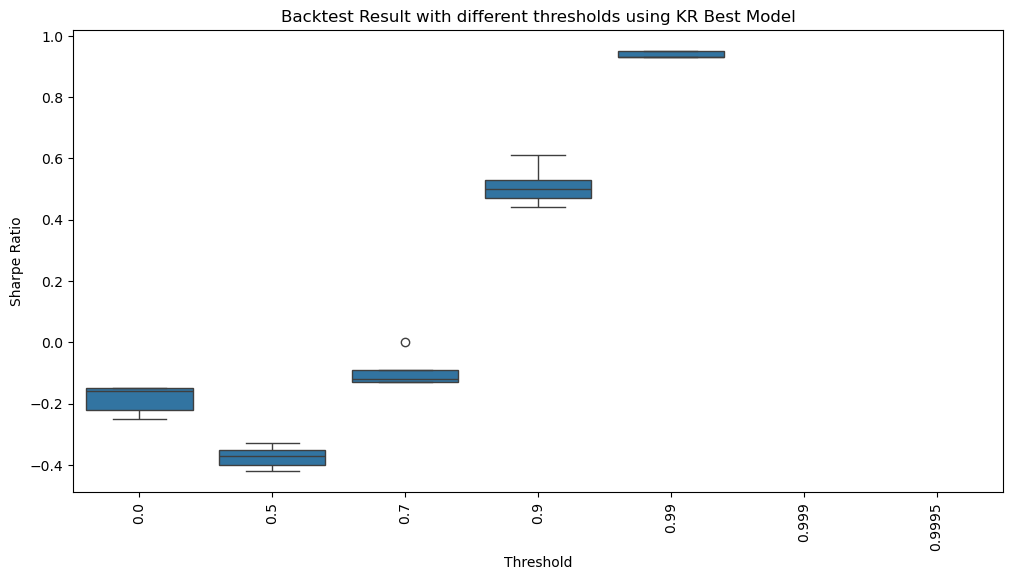}
   \hfil
\caption{Box Plot of Sharpe Ratio with different thresholds during test period using KR Best Model}
\label{kr_threshold_plot_test}
\end{figure}

\begin{figure}[!htbp]
\centering
   \includegraphics[width=11.5cm]{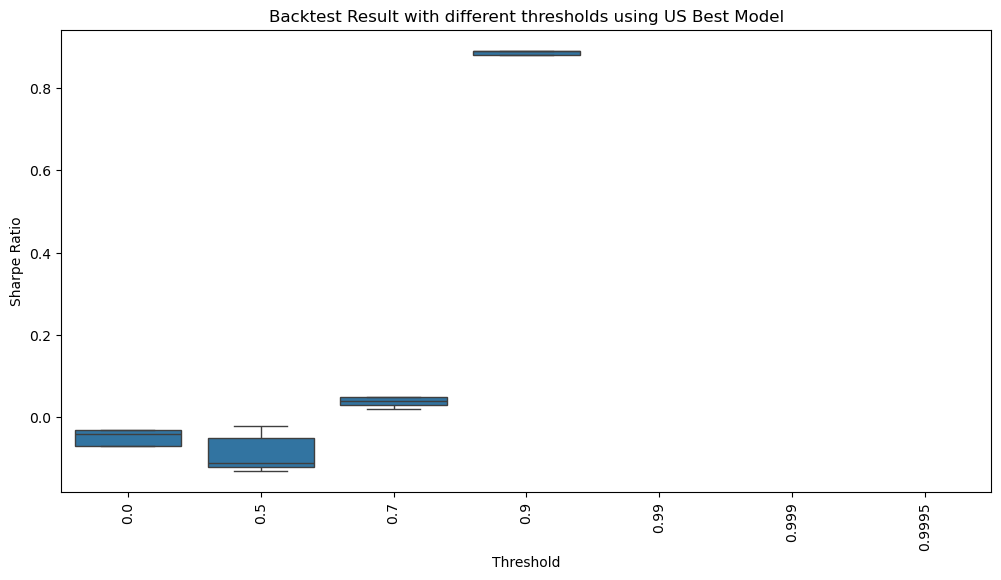}
   \hfil
\caption{Box Plot of Sharpe Ratio with different thresholds during test period using US Best Model}
\label{us_threshold_plot_test}
\end{figure}

The boxplots illustrating Sharpe ratios regarding different thresholds of the best model for KR market is shown on Figure \ref{kr_threshold_plot_test} and and US market on Figure \ref{us_threshold_plot_test}. 

For test period, sharpe ratio tends to increase as you increase threshold, showing a J-like curve. Beyond a certain threshold, the model ceases to provide outputs due to none of the softmax logits surpassing the specified threshold. Because there were big downfall of stock market during the test period, it seems it was advantageous to only buy few stocks that were certain.

\subsection{Example Charts Model Buys}
\begin{figure}[!htbp]
\centering
   \includegraphics[width=15cm]{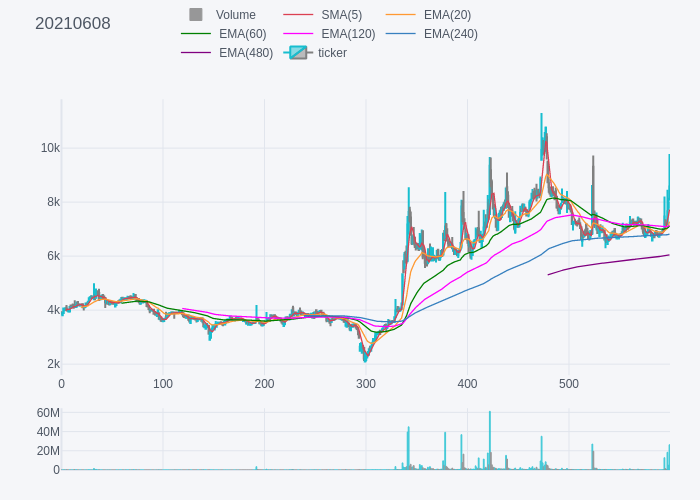}
   \hfil
\caption{A sample of Korea stock test data chart the proposed model will predict 10\% rise}
\label{example chart 1}
\end{figure}

\begin{figure}[!htbp]
\centering
   \includegraphics[width=15cm]{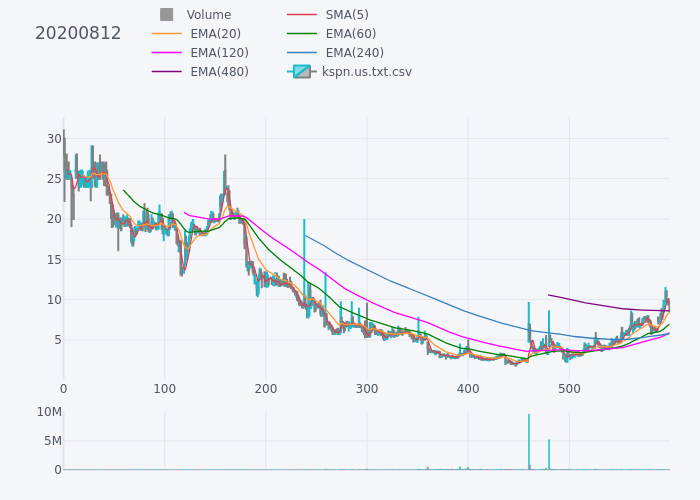}
   \hfil
\caption{A sample of US stock test data chart the proposed model will predict 10\% rise}
\label{example chart 2}
\end{figure}

Stock bar chart of 600 days that model predicts a 10\% rise in price and actually rises are represented in Fig \ref{example chart 1} and \ref{example chart 2}. The Y-axis represents the stock price, measured in Korea Won for Korea stocks and in dollars for US stocks. The X-axis indicates the ordinal position of the input data. The top left corner displays the predicted date. Blue indicates a bullish candlestick and gray indicates a bearish. OHLCV candles are shown with 5-day SMA (Simple Moving Average), 20/60/120/240/480-day EMA (Exponential Moving Average). 

The model buys stocks with various chart patterns, including sideways, uptrend, and downtrend stocks.
\ref{example chart 1} shows an example of a Korea stock that broke through the resistance area of 200-300 days with a long white body candle. It looks like the stock price has risen a lot, but the model expects it to rise additionally, which it does. 
\ref{example chart 2} shows an example of US stock which the price went down in consecutive days while the price-rising trend is formed in the long term. The price bounces off the 480-day moving average and achieves 10\% rise in price in next few days not shown in the chart.
The accuracy of the model varies according to the market condition. Above patterns work well when the market is bullish, however fails rapidly if the market is bearish. One can further consider avoiding these high momentum stocks.

\onecolumn

\section{Discussions and limits}
We notice that although label\_1\_tp10\_ls10 gives the highest F1 macro score for both Korea and Us labels, it doesn't given necessarily gives the best total return. This seems to be do to over-fitting, as there may be no underlying reasons that makes us possible to predict the stocks that will rise 10\% in that day and the model just fits to noise signals. 
Slippage was not taken into account. Slippage in trading refers to the difference between the expected and actual execution prices of a trade, occurring due to market volatility, liquidity, or order size. All the stocks were assumed to be sold at exact 10\% price. This can be a problem for trading low-volume stocks.
Stock gaps were not taken into account, the same fixed profit or loss was taken even when the day's open price was formed above or below the fixed percentage of 10\%. In real-life trading, the investor may get profit/loss more than the specified percentage. 
Delisted tickers were not taken into account. About 2\% to 3\% of Korea and US stocks are delisted annually. This percentage might not be small, but this risk can be reduced by abstaining from trading during the applicable compliance period.
The model was trained using the closing price adjusted after after-hour trades, which means that the model buys stocks at the exact close price at the end of the market. In a real trading scenario, one might have to buy before the market end or place an order the next day at the previous day's close price.


\section{Conclusion}
In this paper, a deep learning model is used to simulate how professional technical analysts trade by looking at the stock charts. Three labels were assigned according to whether a stock increased or decreased by equal to or more than 10\% or 20\% and whether the stock had less than 10\% or 20\% change in price within the given days. A Resnet model with a window size of 600 was used to predict the labels. To increase the probability of the prediction, only stocks that has softmax logit value larger than a chosen threshold were traded.
Backtesting on years 2020 to 2022 shows that the proposed approach in this paper outperforms index benchmarks. On Korea market it gave return of 75.36\% having Sharpe ratio of 1.57, which far exceeds the market return by 36\% and 0.61. On US market it gives total return of 35.54\% with Sharpe ratio of 0.61, which outperforms other benchmarks such as NASDAQ, S\&P500, DOW JONES index by 17.69\% and 0.27, although it fails to outperform AMEX index.
As far as we know, this paper is the first to utilize softmax logits of deep learning models in trading schemes with recent backtest results that far exceeds the market return. 

\section*{Acknowledgements}
This research was supported by Basic Science Research Program through the National Research Foundation of Korea (NRF) funded by the Ministry of Education under support programs (NRF-2016R1D1A1B04933156 and NRF-2019M3E3A1084054), and Brain Korea 21 FOUR Project in 2023.

\bibliography{ref}


\end{document}